\documentclass[letterpaper,english,prl,notitlepage,twocolumn]{revtex4-2}
\usepackage{verbatim}
\usepackage{amsmath}
\usepackage{amssymb}
\usepackage{graphicx}
\usepackage{dsfont}
\usepackage{soul}
\usepackage{mathtools}

\usepackage{hyperref}
\hypersetup{
    colorlinks=true,
    linkcolor=blue,
    filecolor=magenta,      
    urlcolor=cyan,
    pdftitle={Overleaf Example},
    pdfpagemode=FullScreen,
    }

\makeatletter

\usepackage{color}
\usepackage[usenames,dvipsnames,svgnames,table]{xcolor}
\usepackage{graphicx}
\usepackage{comment}
\usepackage{ulem}
\definecolor{violet}{rgb}{0.58, 0.0, 0.83}

\newcommand{\gae}{\lower 2pt \hbox{$\,
\buildrel{\scriptstyle >}\over {\scriptstyle \sim}\,$}}
\newcommand{\lae}{\lower 2pt \hbox{$\,
\buildrel{\scriptstyle <}\over {\scriptstyle \sim}\,$}}

\makeatother

\begin{document}
\title{Floquet Topology Stabilized with Non-Hermitian Driving}
\author{Christopher I. Timms} 
\affiliation{Ph.D., The University of Texas at Dallas, Richardson, TX, USA} 

\begin{abstract}
This study presents a mechanism that enables the stabilization of Floquet systems indefintely; albeit in a manner that allows for noise during each Floquet cycle. This is due to the fact that external qubits are added after each Floquet cycle and these external qubits obtain information about the Floquet system (in this case the Floquet system is the Anomalous Floquet-Anderson Insulator). This information is used to correct the system for noise after which these qubits are removed. The fact that these external qubits are added and then removed after performing operations on the system is what allows for this process to be referred to as non-Hermitian driving. The external qubits effectively act to carry away entropy of the system and therefore allow for the Floquet system to be cooled. In addition, another mechanism is found where the periodic implementation of entanglement for every site of the system at a high frequency during the normal time evolution allows for the system to be highly localized in a manner similar to Anderson localization.
\end{abstract}
\maketitle

\section*{Introduction}

The distinguishing feature of time-periodic Floquet drive is its ability to produce a wide variety of quantum phases of matter that cannot generated in static systems~\cite{Cejnar2019}. These phases of matter include discrete time crystals~\cite{Khemani2016,Keyserlingk2016,Choi2017,Zhang2017,Rovny2018,Else2020}, Floquet symmetry-protected topological states~\cite{Chandran2014,Nathan2015,Potter2016,Po2016,Harper2017,Roy2017,Potter2018,Kolodrubetz2018,Nathan2021}, as well as systems that are described by any number of synthetic spatial and temporal dimensions~\cite{Baum2018,Yang2022,Zhou2016,Hainaut2018,Cai2022,Celi2014}. One issue that plagues the field of Floquet engineering is the occurrence of thermalization that happens as the system undergoes an increasing number of Floquet driving cycles. This thermalization is detrimental to maintaining robust Floquet phases of matter that can persist indefinitely~\cite{Haldar2018,Roy2018,Geraedts2016,Regnault2016}.

This work attempts to resolve some of these issues by presenting a method to bring a series of qubits to a desired target state even if the initial configuration is completely random and unknown. This is done by having two registers, where the qubits being evolved into a desired target state are in register two. Information is transferred about the configuration of qubits in register two to the qubits in register one. This information is used to dictate how to reorient the qubits in register two such that the target state can be achieved. Moreover, this process allows for the entropy of the qubits in register two to be perfectly eliminated in the absence of noise using the information present in register one, while also not increasing the entropy of register one. This means that according to standard quantum theory systems can be set up that temporarily allow for the total entropy to decrease once the appropriate operations have been implemented and that also allow for entropy to be transferred from one location to another, which, in the ideal case, would not result in the generation of any additional entropy. This decrease in the total entropy does not take into account entropy being carried away or being generated by the fields that rotate the qubits or any other process that can reconcile with the laws of thermodynamics.

This work moves onto showcasing how this method can be used for Floquet engineering by employing a Floquet topological system known as the Anomalous Floquet-Anderson Insulator (AFAI), which is constructed using a two-dimensional cylindrical lattice~\cite{Timms2021,Titum2016,Kundu2020}. Here, topological pumping is maintained indefinitely by pushing particles that diffuse to the bottom half of the cylinder back to the top half, which counteracts the process normally associated with the decay of topological current~\cite{Timms2021}. This is done by entangling the particles that make up the AFAI with external qubits and using these entangled qubits to decide how to evolve the system. Finally, an investigation is made into how entanglement operations involving both the particle(s) located on the two-dimensional cylindrical lattice as well as external qubits result in the localization of particles on the lattice.

\section*{Methods}
\subsection*{Basic Setup for Non-Hermitian Driving}
\begin{figure*}
\centering
	\includegraphics[width=0.65\textwidth]{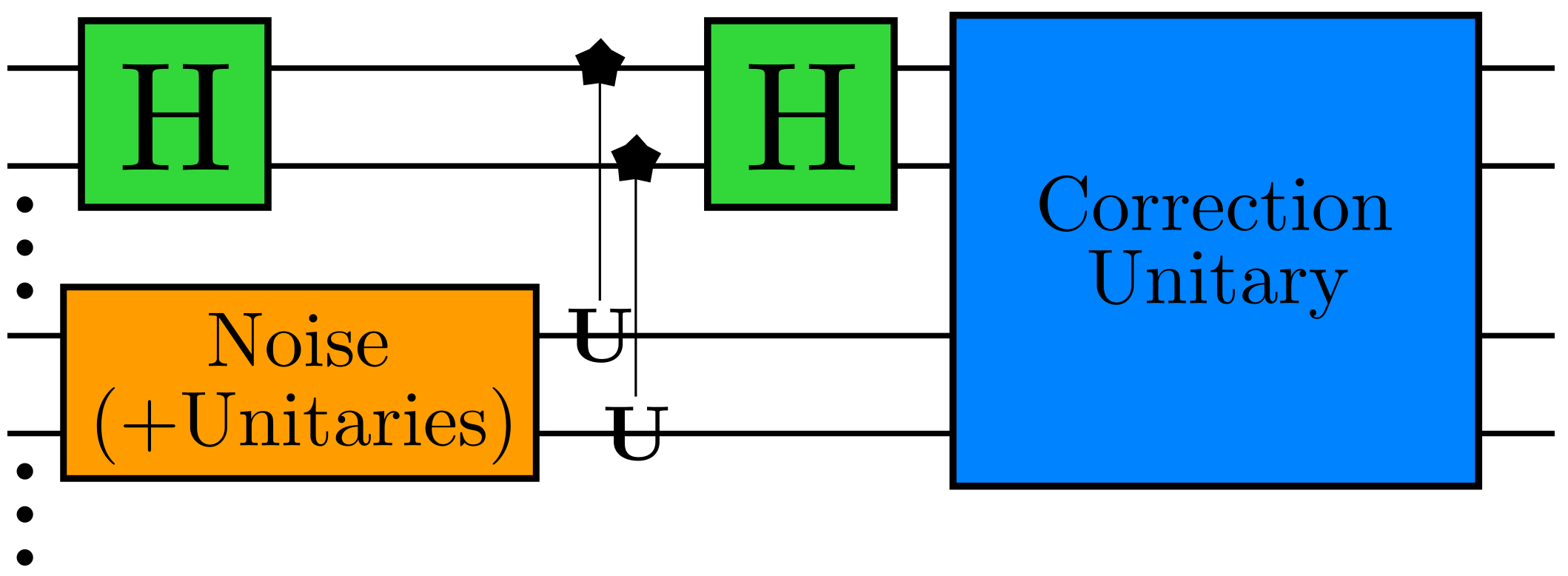}
    \caption{This is an illustration of the basic setup of a quantum circuit which brings any quantum system to a particular target state. There are two registers of qubits, where the qubits in the second register are the ones being evolved into a target state and the qubits in the first register are used to gain information about the qubits in the second register and then reorient these qubits accordingly.}
    \label{CorrectionCircuit}
\end{figure*}

Figure~\ref{CorrectionCircuit} illustrates a quantum circuit that uses qubits in the first register to evolve qubits in the second register into a desired target state. The circuit starts off with a series of Hadamard gates being applied to qubits in the first register, which brings the qubits from the spin down state to a superposition of the spin up and spin down states~\cite{Shepherd2006}. Meanwhile, the qubits in the second register have already been exposed to noise as well as perhaps various unitaries. Within the simulation of this circuit, the noise is implemented through a series of Kraus operators and the particular noise channel that is implemented is the Pauli noise channel. If $\rho_i$ is the initial state of the density matrix that describes the qubit of interest, then the Pauli noise channel is applied as follows:
\begin{equation*}
\rho_f = (1-p_x-p_y-p_z)\rho_i + p_x \sigma_x \rho_i \sigma_x + p_y \sigma_y \rho_i \sigma_y + p_z \sigma_z \rho_i \sigma_z
\end{equation*}
where $\rho_f$ is the final state of the density matrix, $\sigma_x$, $\sigma_y$, and $\sigma_z$ are the x, y, and z Pauli matrices, respectively, and $p_x$, $p_y$, and $p_z$ are the probabilities associated with the rotations using the Pauli matrices~\cite{Chen2023,Chen2022}.

Next, a series of controlled unitaries are implemented where the control qubits are in the first register and the target qubits are in the second register. The controlled unitaries are implemented such that if the first qubit of the first register is the control qubit, then the corresponding target qubit would be the first qubit of the second register or if the second qubit of the first register serves as the control qubit, then the corresponding target qubit would be the second qubit of the second register and so forth. The way that these controlled unitaries are constructed is that there are two kets for each qubit, $|\Psi\rangle_{\mathrm{good}}$ and $|\Psi\rangle_{\mathrm{bad}}$, where $|\Psi\rangle_{\mathrm{good}}$ is the desired target state of the respective qubit and $|\Psi\rangle_{\mathrm{bad}}$ is a state orthogonal to the target state. The controlled unitary would then be set up as follows:
\begin{equation*}
\mathrm{CU} = \begin{bmatrix} 1 & 0 \\ 0 & 0 \end{bmatrix} \otimes I_2 + \begin{bmatrix} 0 & 0 \\ 0 & 1 \end{bmatrix} \otimes (|\Psi\rangle_{\mathrm{good}}\langle \Psi|_{\mathrm{good}}-|\Psi\rangle_{\mathrm{bad}}\langle \Psi|_{\mathrm{bad}})
\end{equation*}
where $I_2$ is the $2\times 2$ identity matrix, $\begin{bmatrix} 1 & 0 \\ 0 & 0 \end{bmatrix}$ is the spin down state, and $\begin{bmatrix} 0 & 0 \\ 0 & 1 \end{bmatrix}$ is the spin up state. If $|\Psi\rangle_{\mathrm{good}} = \begin{bmatrix} 1 \\ 0 \end{bmatrix}$ and $|\Psi\rangle_{\mathrm{bad}} = \begin{bmatrix} 0 \\ 1 \end{bmatrix}$ or vice versa, then the controlled unitary becomes a controlled-z gate. This process applies a phase of $-1$ if the target qubit is in the state orthogonal to the target state and the control qubit is in the spin up state, which will be utilized for interference effects.

After this operation has been completed, another series of Hadamard gates are applied to all of the qubits in the first register. If a particular qubit in the first register ends up in the spin up state after this operation has been completed, then the corresponding qubit in the second register would be in the state $|\Psi\rangle_{\mathrm{bad}}$, whereas if this qubit in the first register ends up in the spin down state, then the corresponding qubit in the second register would be in the $|\Psi\rangle_{\mathrm{good}}$ state. The correction unitary uses this fact to potentially reorient the qubits in the second register, such that these qubits always end up in their respective target state (assuming no noise comes into play). The way that this unitary works is that if a particular qubit in the first register is in the spin down state, then an identity matrix is applied to the corresponding qubit in the second register, whereas if this particular qubit in the first register is in the spin up state, then a rotation matrix defined by $\mathrm{U}_{\mathrm{rot}} = |\Psi\rangle_{\mathrm{good}}\langle \Psi|_{\mathrm{bad}}+|\Psi\rangle_{\mathrm{bad}}\langle \Psi|_{\mathrm{good}}$ is applied to the corresponding qubit in the second register. This unitary can be seen with the following equation:
\begin{equation*}
\mathrm{CU}_{\mathrm{cor}} = \begin{bmatrix} 1 & 0 \\ 0 & 0 \end{bmatrix} \otimes I_2 + \begin{bmatrix} 0 & 0 \\ 0 & 1 \end{bmatrix} \otimes (|\Psi\rangle_{\mathrm{bad}}\langle \Psi|_{\mathrm{good}}+|\Psi\rangle_{\mathrm{good}}\langle \Psi|_{\mathrm{bad}})
\end{equation*}
where $\mathrm{CU}_{\mathrm{cor}}$ is the correction unitary. It is important to keep in mind that $|\Psi\rangle_{\mathrm{good}}$ and $|\Psi\rangle_{\mathrm{bad}}$ can be different for each qubit. An informative example of this process working in practice can be seen on this \href{https://github.com/htim327/NonHermitianDriving/blob/main/docs/source/BasicSetup.rst}{GitHub page}.

\subsection*{Introduction to the Anomalous Floquet-Anderson Insulator}

\begin{figure}[h]
	\includegraphics[width=\columnwidth]{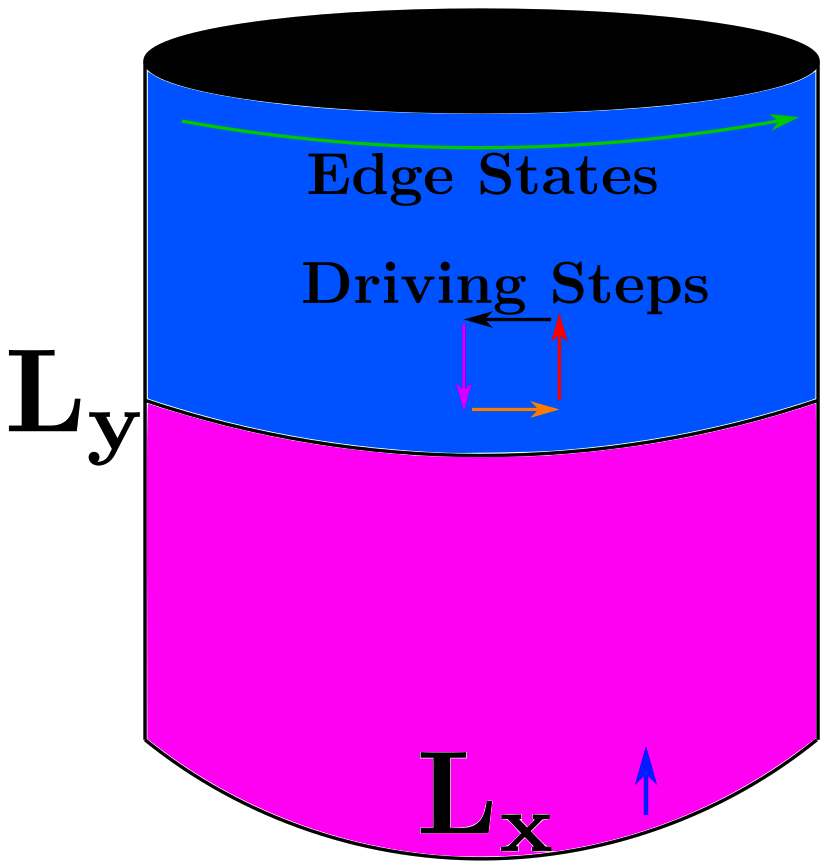}
	\caption{This figure is an illustration of how the Anomalous Floquet-Anderson Insulator (AFAI) is constructed. The top half of the cylindrical two-dimensional lattice (whose dimensions are defined by $\mathrm{L}_\mathrm{x}$ and $\mathrm{L}_\mathrm{y}$ for the x and y directions respectively) are populated with particles as shown in blue, while the pink region is empty of particles. A series of driving steps are periodically implemented and are represented by the colored arrows that form a square. In this case, the arrows represent a particle initialized at the bottom end of the red arrow. In the absence of any kind of noise and during the first driving step, this particle moves up one site as defined by the red arrow. The rest of the arrows depict the particle's path during the other three driving steps and the result is that the particle returns to its original location. The fifth step is for the application of a chemical potential. The presence of boundary conditions at the upper edge of the cylinder causes there to be an edge state current. The blue arrow at the very bottom represents the implementation of the non-Hermitian drive that pushes particles up to the top half.}
	\label{AFAIDepiction}
\end{figure}

Figure~\ref{AFAIDepiction} depicts how the Anomalous Floquet-Anderson Insulator (henceforth referred to as the AFAI) is set up. This is an illustration of a two dimensional cylindrical lattice with the top half of the cylinder populated with particles (blue). This system undergoes a five step Floquet drive where the first four steps are depicted by the colored arrows at the bottom of the region colored in blue and the last step is to implement a chemical potential. The first four steps are implemented using the Hamiltonian $H_n=-J\sum_{<ij>_n} c_i^\dagger c_j$, where $n$ represents the driving step whereas $i$ and $j$ represents the sites on the lattice that the particles hop between. In the figure, if a particle starts out at the bottom of the red arrow and there is no noise, then the particle will circulate such that it returns to its original position. However, due to the fact that this is a Hermitian Hamiltonian, if the particle was initialized at the end of the pointing direction for either the red or purple arrows, but not for the black arrow, then the particle would circulate in the opposite direction. Meanwhile, the fifth step is implemented using the Hamiltonian $H_5=\Delta \sum_{i} \eta_i c_i^\dagger c_i$, where $\eta_i=+1$ ($-1$) such that the opposite potential is implemented for each adjacent site. $J$ and $\Delta$ are chosen such that $J=5\Omega/4$ and $\Delta = 0.4\Omega$ where $\Omega=\hbar=1$. 

Moreover, this two dimensional lattice can be divided into unit cells that are one site wide in the y-direction and two sites long in the x-direction. The left A site is subject to a phase as a result of the chemical potential that is opposite to the right B site. The quantity $\mathrm{L}_x$ can define the number of unit cells for the lattice in the periodic x-direction and $\mathrm{L}_y$ can define the number of unit cells in the y-direction, which for this study is two and four, respectively~\cite{Timms2021,Titum2016,Kundu2020,RodriguezMena2019}.

For the noise, both chemical potential disorder as well as temporal noise are implemented. The chemical potential disorder noise is applied using the Hamiltonian $H_\mathrm{dis}=\sum{\mu_i c_i^\dagger c_i}$, where each $\mu_i$ is uniformly drawn from the interval $[-W,W]$~\cite{Titum2016,Timms2021}. Meanwhile, the temporal noise is implemented by altering the amount of time used for the time evolution associated with each of the driving steps. If, in the absence of temporal noise, each Floquet cycle is evolved over a time of $T=2\pi$ and each step requires a time of $T_n=T/5=2\pi/5$, then the temporal disorder is applied using the equation $T_n = T(1+\delta_n)/5$ with $\delta_n\in [-W_T,W_T]$ being resampled for each Floquet cycle~\cite{Timms2021}.

Finally, the topological quantity that is measured for this system is the charge pumped per Floquet cycle and is given by:
\begin{equation*}
\label{ChargePumpedEquation}
Q = \int_{0}^{\tilde T} dt \langle\psi(t)|I_x|\psi(t)\rangle,
\end{equation*}
where $I_x$ is the current in the x direction and $\tilde T = \sum_n T_n$ is the time taken for each cycle when the temporal noise is taken into account~\cite{Timms2021}. The code that simulates the general setup of the AFAI can be seen on this \href{https://github.com/htim327/NonHermitianDriving/blob/main/docs/source/AFAI.rst}{GitHub page}. This gives a topology that cannot be achieved in a static system and recently, it has been shown numerically that this topology can remain robust to finite levels of the temporal and chemical potential disorder, but becomes non-existent after a certain threshold of both types of disorder have been reached~\cite{Timms2021}. However, the purpose of this work is to show that this topology can be maintained at even higher levels of disorder using the approaches presented and thus imply that a wide variety of Floquet systems, including the AFAI, can be maintained indefinitely in real world environments.

\subsection*{Application of non-Hermitian driving to the AFAI}

Previous studies have shown that in the presence of noise (chemical potential and temporal), the topological current decays over time for many systems, but can remain topologically robust to weak enough levels of both types of noise. This decay is due to the fact that particles diffuse from the top half of the system to the bottom, which creates a counter-propagating current on the bottom half of the cylinder that causes the net current to be reduced to zero~\cite{Timms2021,Titum2016}. Therefore, the main goal that this study set out to accomplish was to push particles from the bottom half of the cylinder back to the top half through the use of non-Hermitian driving. This is done by starting at the very bottom of cylinder and iteratively performing a set of operations, which will be defined later, on each of the sites in the x-direction. After this is done, this process is completed again for each of the sites that are just above those at the very bottom of the cylinder. This is done until the boundary between the top half and the bottom half is reached.

The way that these sets of operations work is that as these operations are iteratively implemented on each of the sites starting at the very bottom of the cylinder, these operations also involve the site just above the site of interest. The site of interest defines where particles are supposed to be pushed away from and the site above the site of interest defines where particles are supposed to be pushed towards. This is represented by the blue arrow in Figure~\ref{AFAIDepiction} where a random site is chosen at the very bottom of the cylinder to implement the non-Hermitian driving protocol that pushes particles away from the bottom. The first step to these operations is to determine whether more particles occupy the top or bottom site and place that information on an external qubit. Unfortunately, this means that instead of simulating the non-interacting particles one at a time as per the normal implementation of the AFAI~\cite{Timms2021,Titum2016,Kundu2020}, it is necessary to simulate all of the relevant particles at the same time, which dramatically increases the computational cost with each additional particle. Therefore, the wave function to describe this system becomes:
\begin{equation*}
\Psi_{\mathrm{tot}} = \begin{bmatrix} \Psi_1 \\ \Psi_2 \\ \Psi_3 \\ . \\ .\\ . \end{bmatrix}
\end{equation*}
where $\Psi_1$ is the wave function of the first particle, $\Psi_2$ is the wave function of the second particle, and so forth. The second step involves using the resulting outcome of the first operation on the external qubit to decide the appropriate operation to apply to $\Psi_{\mathrm{tot}}$. The final operation involves removing the external qubit by finding the appropriate reduced density matrix.

The addition of the external qubit simply involves $\Psi_{\mathrm{tot+ext}} = \Psi_{\mathrm{tot}}\otimes\begin{bmatrix}1 & 0 \\ 0 & 0\end{bmatrix}$. Due to the computational cost of this simulation, the AFAI was only populated with two particles initialized at the very top of the cylinder, but this is advantageous for explaining how the non-Hermitian operations are executed. If the AFAI cylinder is initially populated with two particles, the initial operation that flips the external qubit depending on how many particles are present at each of the two sites involved is represented by the unitary:
\begin{multline*}
\mathrm{CU}_{\mathrm{NH1}} = |\Psi_k\rangle\langle\Psi_k|\otimes|\Psi_k\rangle\langle\Psi_k|\otimes\sigma_x+|\Psi_k\rangle\langle\Psi_k|\otimes I_{\sim k,\sim m}\otimes \sigma_x \\ +I_{\sim k,\sim m}\otimes|\Psi_k\rangle\langle\Psi_k|\otimes \sigma_x+|\Psi_m\rangle\langle\Psi_m|\otimes|\Psi_m\rangle\langle\Psi_m|\otimes I_2+ \\ |\Psi_m\rangle\langle\Psi_m|\otimes |\Psi_k\rangle\langle\Psi_k|\otimes I_2+|\Psi_k\rangle\langle\Psi_k|\otimes |\Psi_m\rangle\langle\Psi_m|\otimes I_2+ \\ |\Psi_m\rangle\langle\Psi_m|\otimes I_{\sim k,\sim m} \otimes I_2+I_{\sim k,\sim m} \otimes |\Psi_m\rangle\langle\Psi_m|\otimes I_2+ \\ I_{\sim k,\sim m}\otimes I_{\sim k,\sim m}\otimes I_2
\end{multline*}
where (for a single particle wave function) $|\Psi_k\rangle$ represents the site where particles are supposed to be pushed away from, $|\Psi_m\rangle$ is the site directly above $|\Psi_k\rangle$ that is supposed to receive particles, $I_{\sim k,\sim m}$ represent all the states that are not $k$ or $m$, and $I_2$ and $\sigma_x$ represent the $2\times 2$ identity and Pauli-x matrices, respectively. It should be noted that, in the real world, one potentially problematic feature of this operation is that it will be desirable to construct an AFAI with larger system size and a larger number of particles~\cite{Zhang2022,Timms2021} and a system involving a wave function composed of many particles will allow more particles to potentially occupy any given site. So either more external qubits need to be added to count the number of particles on each of the sites, which adds more complexity and noise, or a special unitary needs to be constructed that approximates this operation, which potentially reduces the complexity but adds noise. The fortunate news is that it becomes less and less likely for more and more particles to occupy a particular site and so, in a certain sense, there might be a theoretical maximum number of particles that can occupy a particular site, which can substantially decrease the number of qubits needed for this operation. Alternatively, the Pauli Exclusion Principle can be utilized along with fermionic particles to limit the occupation per site to a single particle~\cite{Omran2015}.

The second operation for the non-Hermitian drive, which involves transforming the entire AFAI system depending on the state that the external qubit is in, is given by the following controlled unitary:
\begin{multline*}
\mathrm{CU}_{\mathrm{NH2}} = I_{\mathrm{AFAI}} \otimes I_{\mathrm{AFAI}} \otimes \begin{bmatrix} 1 & 0 \\ 0 & 0\end{bmatrix} + \\ (|\Psi_k\rangle \langle \Psi_m|+|\Psi_m\rangle \langle \Psi_k|)\otimes (|\Psi_k\rangle \langle \Psi_m|+|\Psi_m\rangle \langle \Psi_k|)\otimes \begin{bmatrix} 0 & 0 \\ 0 & 1\end{bmatrix}
\end{multline*}
where $I_{\mathrm{AFAI}}$ is an identity operation that evolves a single particle wave function. The final operation involves the calculation of the reduced density matrix for all of the qubits other than the external qubit.

\subsection*{Applying a noisy non-Hermitian drive}

One of the primary goals of this paper is to construct a modified AFAI that can generate an eternally robust current in substantially noisy environments. Therefore, it is important to incorporate noise models that realistically simulate the types of noise that the AFAI will be subject to during the non-Hermitian driving protocol, in addition to the chemical potential and temporal disorder already implemented for the standalone AFAI. There are essentially two noise models that are applied to the non-Hermitian drive using Kraus operators. Again, if $|\Psi_k\rangle$ represents the site where particles are supposed to be pushed away from and $|\Psi_m\rangle$ represents the site that is supposed recieve the particles, then one noise model would cause mistakes in the transition of particles from state $|\Psi_k\rangle$ such that this transition does not happen and may even cause particles to flow from state $|\Psi_m\rangle$ to state $|\Psi_k\rangle$. This can be done using the Kraus operators $\mathrm{K}_1=I$ and $\mathrm{K}_2=(|\Psi_k\rangle\langle \Psi_m|+|\Psi_m\rangle\langle \Psi_k|+I_{\sim k,\sim m})$, where $I$ is the identity operator and $I_{\sim k,\sim m}$ is the identity operator without sites $k$ or $m$ populated. However, a more accurate noise model would be one where each particle that composes the AFAI is affected by the noise separately, which can be calculated for the AFAI composed of two particles using:
\begin{multline*}
\mathrm{K}_1=I_{\mathrm{AFAI}}\otimes I_{\mathrm{AFAI}} \\ \mathrm{K}_2=I_{\mathrm{AFAI}}\otimes (|\Psi_k\rangle\langle \Psi_m|+|\Psi_m\rangle\langle \Psi_k|+I_{\sim k,\sim m}) \\ \mathrm{K}_3=(|\Psi_k\rangle\langle \Psi_m|+|\Psi_m\rangle\langle \Psi_k|+I_{\sim k,\sim m}\otimes I_\mathrm{AFAI})
\end{multline*}
If $\rho_i$ and $\rho_f$ describe the initial and final density matrix of the system, respectively, then this noise channel would be implemented through $\rho_f = (1-\gamma)\mathrm{K}_1 \rho_i \mathrm{K}_1^\dagger+\gamma \mathrm{K}_2 \rho_i \mathrm{K}_2^\dagger+\mathrm{K}_3 \rho_i \mathrm{K}_3^\dagger$, where $\gamma$ describes the strength of the noise. This noise model is applied immediately after the second non-Hermitian driving step has concluded and the external qubit is separated from the system.

The other noise model mistakenly moves particles from states $|\Psi_k\rangle$ and $|\Psi_m\rangle$ to either one of the four sites that surround either of these two states. If $l=1-4$ is an index that describes the four sites that surround a single particular site, then the corresponding Kraus operators for the two particle system would be given by:
\begin{multline*}
\mathrm{K}_1=I_{\mathrm{AFAI}}\otimes I_{\mathrm{AFAI}} \\ \mathrm{K}_{1+l}=I_{\mathrm{AFAI}}\otimes (|\Psi_k\rangle\langle \Psi_l|+|\Psi_l\rangle\langle \Psi_k|+I_{\sim k,\sim l}) \\ \mathrm{K}_{5+l}=I_{\mathrm{AFAI}}\otimes (|\Psi_m\rangle\langle \Psi_l|+|\Psi_l\rangle\langle \Psi_m|+I_{\sim m,\sim l}) \\ \mathrm{K}_{9+l}= (|\Psi_k\rangle\langle \Psi_l|+|\Psi_l\rangle\langle \Psi_k|+I_{\sim k,\sim l})\otimes I_{\mathrm{AFAI}} \\ \mathrm{K}_{13+l}= (|\Psi_m\rangle\langle \Psi_l|+|\Psi_l\rangle\langle \Psi_m|+I_{\sim m,\sim l}) \otimes I_{\mathrm{AFAI}}
\end{multline*}
Here, $\gamma_2$ would describe the strength of the noise defined by this noise model and this noise model is applied before and after both the non-Hermitian drive and the other noise model are implemented. The code that simulates this noisy application of the non-Hermitian drive on the AFAI can be seen on this \href{https://github.com/htim327/NonHermitianDriving/blob/main/docs/source/ModifiedAFAI.rst}{GitHub page}.

\begin{figure*}
\centering

\includegraphics[width=1\textwidth]{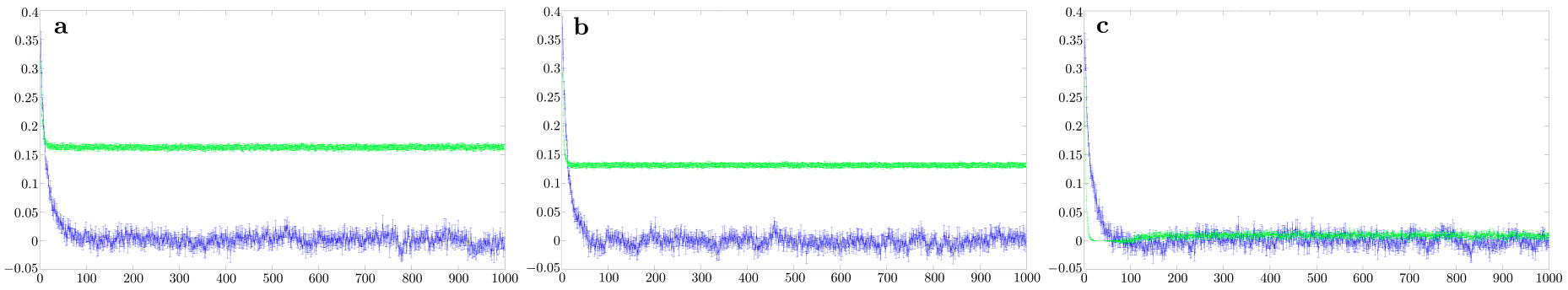}
\caption{These are plots that show how the results differ between the non-Hermitian protocol (green) and the regular setup of the AFAI (blue). The value for $W$, $W_T$, and $\Delta$ are 1.5, 0.2, and 0.4, respectively, for all of these figures. For the noise models associated with the non-Hermitian protocol, a) uses $\gamma=\gamma_2=0.01$ and is supposed to represent what is achievable with current or near-term hardware, b) uses $\gamma=\gamma_2=0.03$ and is supposed to represent what is easily achievable with current hardware, and c) uses $\gamma=\gamma_2=0.06$ and was supposed to be a bragging representation of yesterdays hardware. However, the system may grow more robust as the system size and number of particles grows~\cite{Timms2021}.}
\label{topology}
\end{figure*}

\section*{Results}

\subsection*{Basic Setup for Non-Hermitian Driving}

This \href{https://github.com/htim327/NonHermitianDriving/blob/main/docs/source/BasicSetup.rst}{GitHub page} describes the techniques implemented for the corresponding sub-section within the `Methods' section. Within, it is shown that these techniques do in fact transform the qubits in the second register to the appropriate target states using the qubits in the first register and all of the entropy present in the second register is transferred to the first register. It is also shown that before the correction unitary is implemented, the entropy of both registers is each individually higher than that of the first register after the correction unitary is implemented (with the corresponding entropy of the second register being zero). This makes it seem as though a system can be specifically set up involving two exact copies of a sub-system, each with the exact same entropy, where the entire entropy of the total system can be made to decrease. However, it is entirely conceivable that the fields that implement the correction unitary either carry the missing entropy away or generate new entropy, since the fields can perform multiple actions at once depending on the state of the system of interest.

\subsection*{Enhanced Topology}

The results from Figure~\ref{topology} clearly show an enhanced topology even in the presence of relatively high levels of noise. These results use the values $W=1.5$, $W_T=0.2$, $\Delta=0.4$, and sub-figures (a-c) use $\gamma=\gamma_2$ even though $\gamma$ might be different from $\gamma_2$ in the real world. The topological response decays rather quickly for the case where it is supposed to be enhanced (in green) partially due to the noise of the non-Hermitian protocol, but the topological response then stabilizes rather quickly afterward. This seems to be true for Figures~\ref{topology} a) and b), but decays to being close to zero for c) where $\gamma=\gamma_2=0.06$.

What is remarkable about these results is that if site $k$ and site $m$ are represented as a single qubit and the external qubit is also included to form a two qubit system, then  these sub-figures show that quantum hardware that is either already available or will be developed in the near-term can actually achieve this topological enhancement. Figure~\ref{topology} a) shows what can be achieved with near term hardware, Figure~\ref{topology} b) shows what might be possible to achieve with current quantum hardware, while Figure~\ref{topology} c) is supposed to show what can definitely be achieved with quantum hardware that is easily available~\cite{Gullans2019,Kim2022,Zahedinejad2016,Moskalenko2022,Heinz2021,Huang2019,Xue2019,Wang2020,Xu2020}. The designations as near term hardware, currently available hardware, and easily available hardware are supposed to be rather conservative, so it could be possible that systems in the real world will outperform these results. It is important to remember that a unitary probably needs to be fitted for this operation if the two sites as well as the external qubit are to be represented as two qubits (which will probably involve the imposition of some errors), otherwise more qubits will likely be needed for this process to occur so that the total number of particles per site can be counted. However, as stated before, fermionic particles can utilize the Pauli Exclusion Principle to limit the maximum number of particle per site to one~\cite{Omran2015}, which would dramatically reduce the complexity of the problem.

\subsection*{Localization Induced by Quantum Entanglement}
\begin{figure}[h]
	\includegraphics[width=\columnwidth]{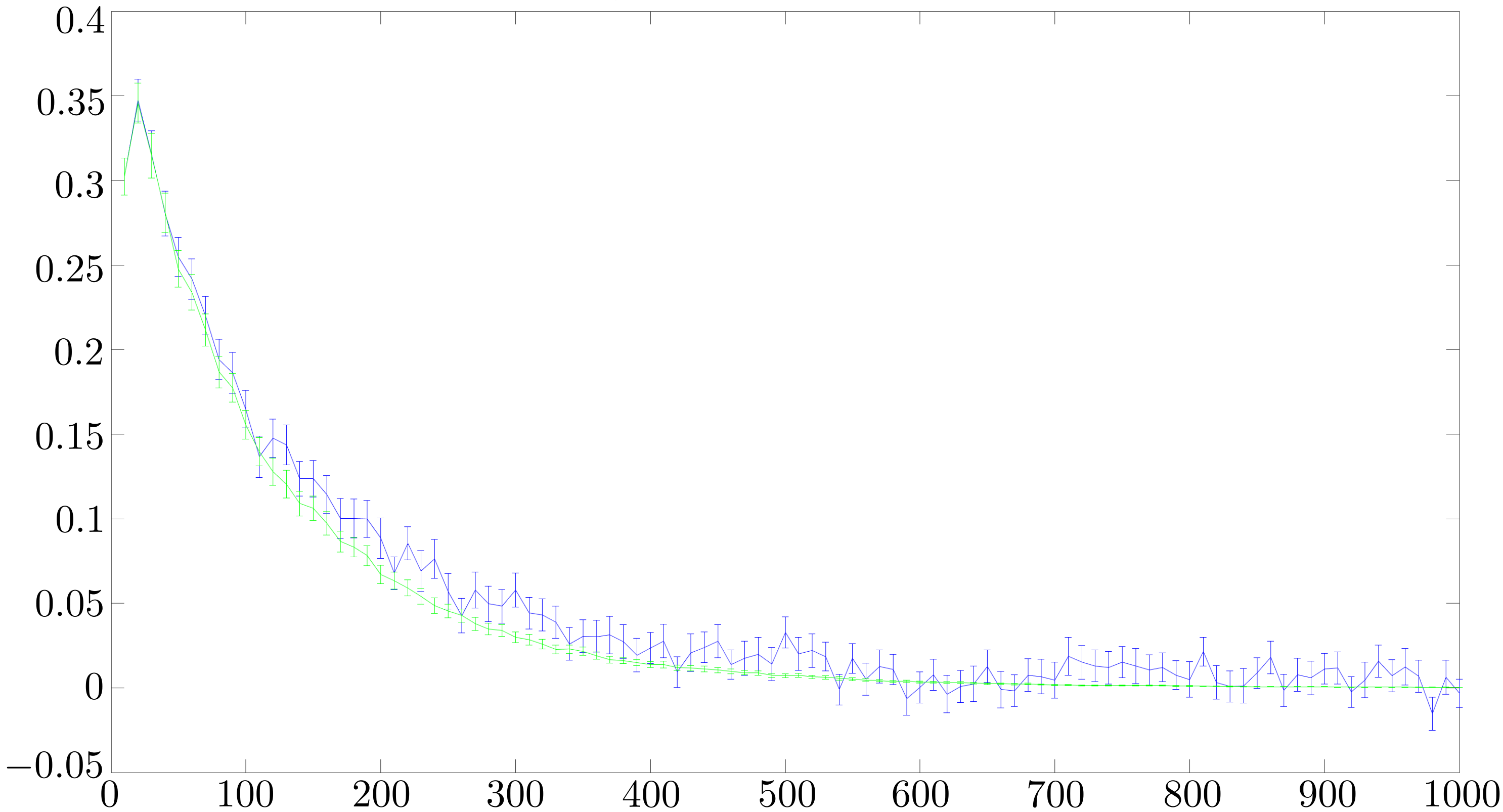}
	\caption{This depiction is a check to find software bugs in the code that is designed to simulate the AFAI subject to the non-Hermitian drive. This plot uses $W=1.5$, $W_T=0.2$, $\Delta=0.4$, and zero noise for the implentation of the non-Hermitian drive. In addition, the system is not subject to the correction unitary, but the presence of particles at various sites is entangled with an external qubit. The initial expectation was that the curve that includes the external qubit (green) should exactly match the curve that does not include this qubit (blue). The reason this does not happen is argued within this sub-section.}
	\label{ZeroDisorderTest}
\end{figure}

Figure~\ref{ZeroDisorderTest} was originally a check to find software bugs in the code that simulates the AFAI being subject to the non-Hermitian drive. This uses the same setup as the subsection above, but zero noise is used for the implementation of the non-Hermitian drive and the curve that includes an external qubit (green) is not subject to the correction unitary. It was initially expected that the two curves should match exactly due to the fact that this operation does not impact the probability of the particles occupying each of the sites, but it was later realized that this operation does change the phase of the particles that occupy any given site as is seen in this \href{https://github.com/htim327/NonHermitianDriving/blob/main/docs/source/BasicSetup.rst}{GitHub repository}. This process of having a phase change that is highly dependent on the site of interest looks to be very similar to the implementation of chemical potential disorder~\cite{Timms2021}, which partially explains why the topological response of the system that involves the external qubit decays faster than the original setup of the AFAI.

\begin{figure}[h]
	\includegraphics[width=\columnwidth]{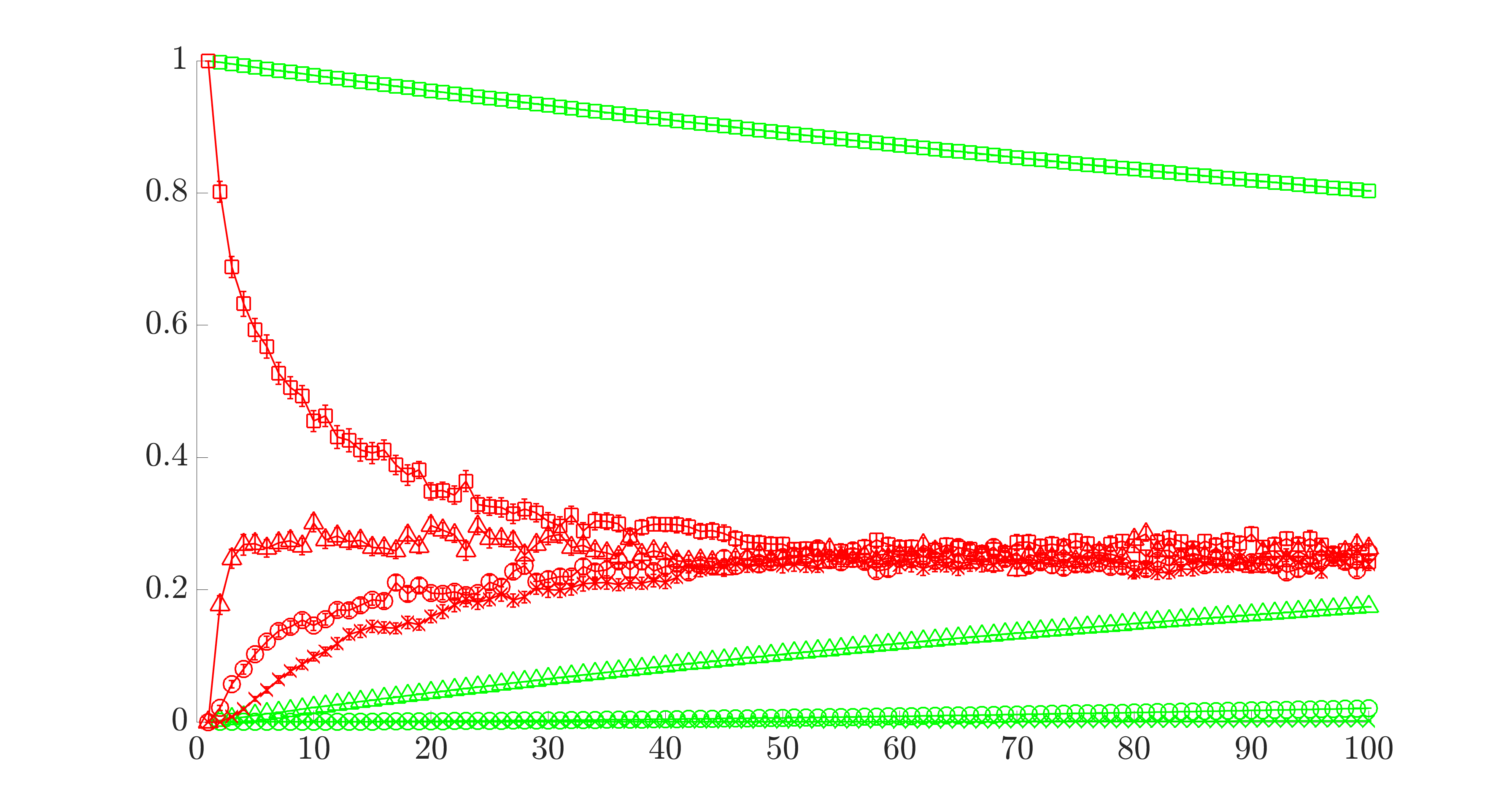}
	\caption{This illustration shows the results when the unitaries that define the driving steps are divided by 1000 and information about where the single particle is located is entangled with an external qubit. The entanglement process is completed for every site of the system with an external qubit being added and removed for each site. As with the previous plot, the correction unitary is not implemented. This plot uses $W=1.5$, $W_T=0.2$, $\Delta=0.4$, and zero noise for the implementation of the non-Hermitian drive. The squares have a y-index of 0, the triangles have a y-index of 1, the circles have a y-index of 2, and the crosses have a y-index of 3. The green curve represents the system where localization is induced and the red curve represents the system where the quantum entanglement is not generated to induce localization.}
	\label{EntanglementLocalization}
\end{figure}

To test this hypothesis, a setup (seen in this \href{https://github.com/htim327/NonHermitianDriving/blob/main/docs/source/LocalizationByEntanglement.rst}{GitHub repository}) is constructed where the normal driving steps are used, zero disorder is used for the entanglement operations involving the external qubits, and the correction unitary is not implemented. The difference between this setup and the previous one is that the time evolution of each of the driving steps is divided by a factor of one thousand and after the application of each of these unitaries, an iterative process that passes over each of the sites individually is used where an external qubit is added, the presence of the single particle at the site of interest is entangled with this external qubit, and then this external qubit is removed. Therefore, the driving steps are still completed, but the difference is the high levels of entanglement. 

Due to the fact that such little time evolution takes place between the application of on-site quantum phase changes that result from the quantum entanglement, this process should be very similar to Anderson localization because the actual on-site quantum phase changes for each of the sites will not change very much with time after each of the unitaries are implemented~\cite{Skipetrov2008,Billy2008,Timms2021}. Figure~\ref{EntanglementLocalization} confirms this with the green curves illustrating the process that involves entanglement and the red curves illustrating the normal setup of the AFAI. The squares represent a y-index of 0, where the particle is initialized, the triangles represent a y-index of 1, the circles represent a y-index of 2, and the crosses represent a y-index of 3. The red curves show that the particle diffuses away from the y-index of 0 until there is a uniform probability for the particle to occupy each of the y-indices. The green curves show that the particle is much more strongly localized at the y-index of zero and will take a much longer time to reach equilibrium, which is consistent with the process of Anderson localization~\cite{Skipetrov2008,Billy2008,Timms2021}.

\section*{Discussion}

This work has provided a theoretical means for stabilizing a Floquet topological current indefinitely when it would otherwise decay to a value of zero. The intention was that this work would provide an ample amount of luxury for guiding the construction of a system that can be achieved in the real world. Moreover, this represents a system that should theoretically be scalable to any system size (barring longer execution times associated with the implementation of the non-Hermitian drive for larger systems). This should provide good practice for learning how to connect more qubits together, learning how to build bigger quantum simulators, or learning how to reduce noise levels for systems that involve a large number of qubits.

This problem shows how quantum computing technology can already be useful and even be potentially used to obtain quantum advantage or supremacy~\cite{Preskill2012,Preskill2018,Zhong2020}. A quantum computer composed of only a few qubits can be interfaced with a quantum simulator or other type of quantum device to, ideally, dramatically increase abilities with respect to achieving Floquet topology as well as producing stable versions of other Floquet systems. Although this would involve constructing a new small quantum computer for every site that is iterated over, which admittedly could cause problems with respect to time scales involved. The fact that this problem scales dramatically on a classical computer and requires nothing more specific than the simple enhancement of the topological response as well as relatively simple quantum operations makes this problem ideal for claiming quantum advantage or supremacy. If longer execution times associated with the implementation of the non-Hermitian drive do not prevent scaling to an `arbitrary' number of qubits, then perhaps a new term such as `quantum ascendancy' should classify this problem since the doors would be opened to using truly large systems for quantum computation.

In addition, it was found by fortuity that when entanglement entropy is periodically produced at a high frequency for every site of the system, a high degree of localization occurs due to the phases that are added to each of the sites, which is a phenomenon similar to Anderson localization~\cite{Skipetrov2008,Billy2008,Timms2021}. It is unclear what would happen to the degree of localization if the entanglement were implemented at non-periodic instead of periodic intervals, but it is suspected that this will cause a higher degree of thermalization~\cite{Rieder2018,Timms2021}. This behavior may provide more information about entanglement as well as the behavior of density matrices associated with systems subjected to interactions with the external environment~\cite{Nikolic2005,Zurek2003,Zurek2000,Ollivier2001}.

\section*{Acknowledgements}

The author is grateful for the educational resources provided by The University of Texas at Dallas as well as those provided by the team at Amazon Braket.

\bibliographystyle{plain}
\bibliography{NonHermitian}

\end{document}